Common basis for cellular motility

Henry G. Zot[1*], Javier E. Hasbun[2], Nguyen Van Minh[3]

*Footnotes:*
[1]Department of Biology, University of West Georgia, Carrollton, GA 30118

[2]Department of Physics, University of West Georgia, Carrollton, GA 30118

[3]Department of Mathematics, University of Arkansas at Little Rock, Little Rock, AR 72204

[*]Correspondance: zot@westga.edu

## ABSTRACT

Motility is characteristic of life, but a common basis for movement has remained to be identified. Diverse systems in motion shift between two states depending on interactions that turnover at the rate of an applied cycle of force. Although one phase of the force cycle terminates the decay of the most recent state, continuation of the cycle of force regenerates the original decay process in a recursive cycle. By completing a cycle, kinetic energy is transformed into probability of sustaining the most recent state and the system gains a frame of reference for discrete transitions having static rather than time-dependent probability. The probability of completing a recursive cycle is computed with a Markov chain comprised of two equilibrium states and a kinetic intermediate. Given rate constants for the reactions, a random walk reproduces bias and recurrence times of walking motor molecules and bacterial flagellar switching with unrivaled fidelity.

## INTRODUCTION

The time a bacterial flagellar motor rotates in the counterclockwise (CCW) and clockwise (CW) direction is a random variable that depends on ligand binding (1). If the regulation were owing to the decay of a simple equilibrium switch, time intervals of CCW and CW rotation (recurrence times) would be exponentially distributed (1-3), i.e. having a decreasing failure rate (DFR, 4). However, recurrence times are found to have a gamma distribution (1), which exhibits increasing failure rate (IFR, 4). IFR suggests that each switch is a sequence of hidden Markov steps (1-3).

IFR distributed recurrence times between linear displacements of single motor molecules, myosin V (MyoV) or dynein, walking on a polymer have also been observed but not explained (5, 6). Each molecule with two mechanical hands binds to a polymer of interacting sites by mass action (7). During a random recurrence time, the polymer-bound molecule moves by detaching and reattaching one hand in a recursive cycle (8). Detachment waits for the catalytic cycle of the lead hand to generate a force on the follower hand (9), and reattachment waits for a favorable catalytic intermediate and spatial orientation of the follower hand (6, 8). A sequence of two waiting times is repeated during a run that terminates with the dissociation of the motor molecule from the polymer. Lifetimes of runs are random variables, which are DFR distributed (5, 6). We hypothesize a causal relationship between IFR distributed recurrence times and motion.

## MATERIALS AND METHODS

### Derivation of summary reactions

Let [*M*] and [*C*] represent molar concentrations of chemical states and [*T*] represent the sum of [*M*] and [*C*]. Then, the ratios [*M*]/[*T*] and [*C*]/[*T*] are the mole fractions, *M* and *C*, respectively.

Formation of *M* from *C* by mass action is a spontaneous process, $C + nU_1 \xrightarrow{\beta k_1^n} M$, for $\beta$ ensemble ($n \geq 1$) interactions ($U_1$). An external process ($\tau_1$) isolates kinetic intermediate (*m*) from $U_1$ by isometric transfer of kinetic energy (Fig. 1). Arrival of a second external process ($\tau_2$), restores the conditions for interactions ($U_2$) that can regenerate *M* (Fig. 1). Odds of completing the cycle of kinetic work are given by $\alpha = k_2/k_-$ as expressed by $M + nU_2 \xrightarrow{\alpha^n} M + nU_1$ (Fig. 2). *M* forms from *C* by nonspontaneous processes, $C + nU_2 \xrightarrow{\beta k_1^n \alpha^n} M$ (Fig. 2). Formation of *M* by the combination of spontaneous and nonspontaneous processes is given by

$$C + nU \xrightarrow{\beta k_1^n} M \tag{1}$$

where $U = U_1 + \alpha U_2$.

*M* decays to *C* by separate pathways, namely, $M \xrightarrow{k_{-1}} nU_1 + C$ and $M \xrightarrow{k_k_{-2}} nU_1 + C$ (Fig. 2). The combination of pathways is expressed by

$$M \xrightarrow{k_{-1} + k_k_{-2}} nU_1 + C \tag{2}$$

**Derivation of system probabilities**

Let $A_i$ and $B_i$ represent probabilities of *M* formation and decay, respectively, for spontaneous ($i$ = 1) and nonspontaneous ($i$ = 2) processes in the system (Fig. 2). Let, *Z* represent the overall rate. Then,

$$A_1 Z = \beta k_1 U_1 U^{n-1} \tag{3}$$

$$A_2 Z = \beta k_1 \alpha U_2 U^{n-1} \tag{4}$$

$$B_1 Z = k_{-1} \tag{5}$$

$$B_2 Z = k_k_{-2} \tag{6}$$

where $\sum_{i=1}^{2} A_i + B_i = 1$.

**Derivation of *M* bias**

*M* bias has the value of the probability of forming *M*, which is given by $A_1 + A_2$. Because chemical work requires a change in mass and mass is constant for kinetic work (Fig. 1), $U_1 = 1 - M$, and $U_2 = M$, respectively. By substituting for $U_1$ and $U_2$ in Eq. 3 and Eq. 4, *M* bias is expressed by

$$M = K_0 (1 - M)(1 + (\alpha - 1)M)^n \tag{7}$$

where $K_0 = \beta k_1^n / (k_{-1} + k_k_{-2})$.

**Regulation by ligand binding**

To account for regulation by ligand, let $[u_i]$ and $[uL_i]$, represent the molar concentrations of binding sites for ligand-free and ligand-bound components, respectively. Then, $u_i + L \overset{K_L}{\rightleftharpoons} uL_i$ describes the binding of ligand ($L$) by mass action given equilibrium constant ($K_L$). From the equilibrium expression, $[uL_i] = K_L[L][u_i]$, normalizing with $[T]$ gives $U_i' = K_L[L]u_i'$, where $U_i' = [uL_i]/[T]$ and $u_i' = [u_i]/[T]$. To account for interactions with ligand-bound components, $U = U_1' + \alpha U_2'$ in Eq. 7, which modifies the expressions of *M* bias thusly,

$$M = K_0' (1 - M)(1 + (\alpha - 1)M)^n \tag{8}$$

where $K_0' = \beta (k_1 K_L [L])^n / (k_{-1} + k_k_{-2})$.

**RESULTS**

As illustrated by the polymer-associated state of the one-handed motor molecule, a displacement may strain elastic bonds to the point of yielding but a change in chemical state need not occur. Let *C* and *M* represent the chemical states of a protein structure and let $U_1$ represent spontaneous interactions that shift the equilibrium in favor of *M* (*M* bias) by mass action. *M* decays as a continuous function of time unless the arrival of an external process ($\tau_1$) displaces and isolates a kinetic intermediate (*m*) from $U_1$. Energetically, the plane of kinetic work is orthogonal to the plane of chemical work (Fig. 1). As a result, the displacement can accelerate the process of entropic decay (10) or trap *m* on the kinetic plane. By the arrival of a second external process ($\tau_2$), *m* can gain access to non-spontaneous interactions ($U_2$) that restore *M*. Hence, kinetic work regenerates the spontaneous decay of *M* in a recursive cycle consisting of two sequential steps. The conditional probability of the second step explains how recurrence times can become IFR distributed. Although processes, $\tau_1$ and $\tau_2$, have random lifetimes in continuous time, the steps of the recursive cycle are fundamentally discrete.

A second chance mechanism (*SCM*) (cf. 11, 12) accounts for the mole fractions of any number of components acting independently or in concert to couple mass action with regeneration by a recursive cycle (Fig. 2). The average flow of mass in a system composed of $\beta$ ensembles of *n* components is the balance of competing pathways of *M* formation and decay (Eq. 1, 2). From Eq. 1, $U_2$ is seen to arise only for $\alpha = k_2/k_- > 0$, which is the boundary condition for kinetic work (Fig. 1) and probability of a dynamic recursive cycle, i.e. $A_2 > 0$ (Eq. 4). Hence, external work increases the chance of regeneration and, thereby, sustains the most recent state. In addition, with $A_2 > 0$, the system gains a frame of reference for discrete time. Unlike the continuous time transitions of the spontaneous process, discrete time transitions have time-independent probability (static) determined by relative rates of transition (Fig. 2) given by the state space (Fig. 1). Hence, the state space functions as an autonomous system with emergent properties.

Each step of a dynamic recursive cycle corresponds to steps of a discrete Markov chain. Given static transitional probabilities (Fig. 2), the probability distribution of *C*, *M*, or *m* at the end of a step is given by Markov matrix **J**, $\mathbf{J} = \begin{bmatrix} B_1 & 1-B_1 & 0 \\ B_1 & 0 & 1-B_1 \\ B_1+B_2 & A_2 & A_1 \end{bmatrix}$, where the conditional probabilities $P(C|C)$, $P(M|M)$, and $P(m|m)$ are along the diagonal and probabilities sum to unity across each row. Row 1 represents a component in *C* that remains in equilibrium with *M* until the $\tau_1$ process arrives. Afterward, in row 2, a component in *M* forms *m* with unit probability. For row 3, a component in *m* either waits for the arrival of a $\tau_2$ process or decays to *C* with probability $B_1 + B_2$. When $\tau_2$ arrives, the probabilities that *M* regenerates or *m* remains unchanged are $A_2$ and $A_1$, respectively. From matrix **J,** mean discrete recurrence times can be derived for any number of steps (Supporting Material). By including waiting times, $\tau_1$ and $\tau_2$, matrix **J** can be converted to a continuous time Markov matrix **Q** (Supporting Material), from which mean elapsed recurrence times can be calculated. Alternatively, $\tau_1$ and $\tau_2$ can be used to generate random waiting times for steps of a random walk of *SCM* (Fig. S1).

Important insight is gained by comparing recurrence times under two limiting conditions that differ in the pathway by which *M* decays to *C*, namely, Type 1 (T1) and Type 2 (T2) systems. A T1 system is characterized by decay only through *m*, i.e. $B_1 << B_2$, whereas a T2 system is characterized by decay from *M* only, i.e. $B_2 << B_1$ (Fig. 2). Recurrence times calculated for a given $K_0$ (Eq. 7) using stationary probabilities of matrix **J** (Supporting Material) are plotted for comparison (Fig. 3). To demonstrate consistency, recurrence times determined by random walk (Fig. S1) are plotted with calculated recurrence times (Fig. 3).

Recurrence times of a T1 system diverge from three Markov steps as $K_0$ is increased (Fig. 3A). For $K_0 > 1$, the recurrence times of *M* and *m* approach a minimum of two, corresponding to the steps of the dynamic recursive cycle, and recurrence times of *C* grow larger. If *C* were to represent the unbound state, detachment by a walking motor molecule would become less favorable as $K_0$ increases.

Given a T2 system, the minimum recurrence time is one step for either *C* or the combination of *M* or *m* (*Mm*) as a function of decreasing and increasing $K_0$, respectively (Fig. 3B). The symmetry about $K_0 = 1$ favors separate functions ascribed to *C* and *Mm* and regulation of $K_0$ by ligand. Such would be expected for the distinct states of the bacterial flagellar motor that give rise to opposing directions of rotation.

Single motor molecules, MyoV and dynein, moving in a stepwise manner on a polymer of binding partners each represent a single component system ($\beta = 1$, $n = 1$). Spontaneous turnover of polymer attachment competes with regeneration through a kinetically isolated hand (Fig. 4A). Intervals between displacements (dwell times) and between polymer attachment and detachment (run times) observed for MyoV and dynein (5, 13) are modeled by a continuous time random walk (Fig. S1) using constant transition probabilities (Table 1) and varying the lifetimes of each step ($\tau_1$ and $\tau_2$). From the same random walk, recurrence of *Mm* matches the expected IFR distributed dwell times (Figs. 4B-D) and recurrence of *C* reproduces DFR distributed run times (insets Figs. 4B-D).

Varying the parameters that determine recurrence times of *C*, i.e. $\alpha$ and $K_0$ (Fig. 3A), has little effect on the density of *M* recurrence time distributions (Table 1, Fig. 4B-C). The same waiting time for both processes, i.e. $\tau_1 = \tau_2$, gives reasonable fits of MyoV distributions (Table 1; Fig. 4B-C). Thus, dwell time distributions recorded for MyoV in slow and fast regimes, i.e. 2 μM (Fig. 4B) and 2 mM ATP (Fig. 4C), respectively, are mainly a function of $\tau_1$. However, the dwell times of dynein can only be fit by *M* recurrence times generated with longer waiting times for the $\tau_2$ process, i.e. $\tau_2 > \tau_1$ (Fig. 4D; Table 1).

To switch direction of bacterial flagella rotation, ligand concentration regulates the angle of force applied by multiple independent motor units to a common rotor (14, 15). The odds of a switch depends on ligand occupancy of the subunit delivered to each motor unit (Fig. 5A). Given a recursive cycle of regeneration, CCW and CW rotation are, respectively, assigned to *C* and *Mm* given T2 conditions ($k_k_{-2} = 0$, Table 1). In this regime, a motor unit can only switch direction of torque while interacting with a subunit, otherwise the direction is held constant between interactions (inset, Fig. 5B). By extension, one would expect dynamic recursive cycles for both *C* and *M* formation, i.e. *Cc* and *Mm* (inset Fig. 5C). Although two thermodynamic states and four Markov states reconcile proposals for switching by two states (16) and four states (1, 17), respectively, three Markov states adequately fit extant IFR distributed data. Hence, *SCM* with four Markov states is not considered further.

The experimental data are probability density distributions of times that flagella travel in the CCW and CW direction (dwell times) while bias in the CW direction is held constant (1). For a given *M* bias (Eq. 8), $K'_0$ (Table 1) determines the probabilities (Eq. 3-6) used in the random walk (Fig. S1). The rate of the rotor ($\tau_1^{-1}$, Table 1) is the only adjustable parameter.

*M* bias varies with *n* (inset Fig. 5D; c.f. 12) while ligand concentration is held constant, because, as derived here, *M* bias expresses non-allosteric cooperativity based on simple mass action ligand binding (Eq. 8). This is a concerted response that requires agreement of all *n* components. The response element of the ensemble could be an internal structural feature of a motor unit that prevents switching against a prevailing torque (12). Consequently, each step of the recursive cycle must wait for the longest of *n* random lifetimes to switch the angle of force. Systematic selection of the longest waiting time reduces the variance of recurrence times as *n* increases and gives a better fit of dwell time distributions (Fig. 5B-D). To compensate for longer waiting times, rotation rate of the rotor is increased, i.e. compare $\tau_1^{-1}$ for $n = 10$ and $n = 2$ (Table 1). A recursive cycle linked to torque requires $n \geq 2$.

## DISCUSSION

Long dwell times and non-uniform strides of dynein (13) are consistent with a long waiting time for the regeneration step, as expressed by $\tau_2 > \tau_1$. During the $\tau_2$ process, the follower hand returns to a reactive chemical state and spatial location for interaction with the polymer. But, the probability of regeneration depends on the odds of completing a recursive cycle, $\alpha$, relative to the rate of dissociation of the lead hand, $k_{-2}$. A gating mechanism (9) can act to reduce $k_{-2}$, which has the effect of increasing the probability of longer runs. At the moment of regeneration, a new spontaneous process begins without memory and, hence, free of coordination between hands (6, 13).

A characteristic rapid first step that terminates equilibrium binding of MyoV and dynein (5, 6) is captured in *SCM* as a short $\tau_1$ process resulting in T1 conditions ($k_{-1} = 0$; Table 1). Old and regenerated hands may have distinct phases of the same catalytic cycle, which could bias displacement of alternating hands during the $\tau_1$ process. A spontaneous decay of catalytic intermediates accounts for less common random variables such as reverse and shuffle steps by the same or alternating hand (6, 13).

A plot of CW bias of bacterial flagella rotation is a logistic function of ligand concentration and demonstrates cooperation in the switching mechanism (18). Cooperative *M* bias (inset Fig. 5D) as a function of simple mass action binding of ligand (Eq. 8) is consistent with evidence of one class of ligand binding sites of the rotor (19, 20). Static probability resulting from a recursive cycle gives rise to spatial and temporal precision with respect to ligand bound rotor subunits (Fig. 5A). By contrast, a most favorable form of allosteric regulation predicts two classes of binding sites and propagation of structural changes from remote ligand binding sites to sites of motor units (21, 22).

From the inset Fig. 5D, adding motor units, *n*, can be seen to reduce or increase *M* bias depending on a fixed ligand concentration below or above the midpoint, respectively. Indeed, the recruitment of motor units to a rotor under load is seen to stabilize CCW or CW rotation under conditions that hold ligand concentration constant (23, 24). A change in bias by an allosteric mechanism requires a change in ligand concentration (21, 22).

That the CCW and CW dwell times are DFR distributed for flagella in near zero load conditions is thought to be irreconcilable with IFR distributed dwell times measured under moderate load conditions (25). However, the absence of torque may disconnect spatially disjoined motor units of an ensemble. The absence of torque also reduces the number of motor units per rotor to near unity (23). Both the lack of torque and sufficient ensemble size eliminate waiting times associated with a switch, which restricts *SCM* to spontaneous events having DFR distributed recurrence times. That the dwell times of flagellar stubs under load are also DFR distributed (25) refers to an analysis that discriminates against the shortest reversal events, termed incomplete switches (22). The record includes multiple switches judged incomplete by being too brief to allow the flagellar stub to attain full speed (22). However, the response of the stub to a switch could be damped by elastic elements under load. Indeed, durations of incomplete switches are IFR distributed random variables (22). Therefore, the load necessary for kinetic work can reconcile IFR and DFR distributed recurrence times of flagellar switching.

Motion can be seen as instrumental in regenerating the most recent state by a two-step recursive cycle. A recursive cycle is a prominent feature of processive motors, sliding filaments, and rolling cell adhesion. Two consecutive waiting times separate angular displacements of the F1-ATPase during a catalytic cycle (26). Cooperative ligand dependency of the most recent state of bacterial flagellar switching (12) and muscle activation (11) are described by an *SCM*-derived non-allosteric relationship (Eq. 8). Hence, SCM can explain the behavior of diverse forms cellular motility.

We see here that a time-continuous process of decay becomes discrete when the kinetic intermediate jumps back to regenerate the most recent state. The system is open, but the state space functions as an autonomous system that preserves detailed balance as interactions turn over and evolves in time as a discrete Markov chain. Units of *SCM* can be regulated, modulated, and combined in parallel or in series to provide stability, binary output, non-allosteric cooperativity, and fine spatial and temporal control. These emergent properties are selectable traits, which become the basis for diverse forms of motility over time. Further study of discrete properties may shed light on how complexity may be built through natural selection of *SCM* units.

**REFERENCES**

1. Korobkova, E. A., T. Emonet, H. Park, and P. Cluzel. 2006. Hidden stochastic nature of a single bacterial motor. *Phys. Rev. Lett.* 96:058105.
2. Tu, Y. 2008. The nonequilibrium mechanism for ultrasensitivity in a biological switch: sensing by Maxwell's demons. *Proc. Natl. Acad. Sci. USA.* 105:11737–11741.
3. van Albada, S. B., S. Tănase-Nicola, and P. R. ten Wolde. (2009) The switching dynamics of the bacterial flagellar motor. *Mol. Syst. Biol.* 5:316.
4. Ross, S. M. 2010. System life as a function of component lives. *In* Introduction to Probability Models, ed. 10. Elsevier, New York, pp. 602-610.
5. Rief, M., R. S. Rock, A. D. Mehta, M. S. Mooseker, R. E. Cheney, and J. A. Spudich. 2000. Myosin-V stepping kinetics: a molecular model for processivity. *Proc. Natl. Acad. Sci. USA.* 97:9482-9486.
6. Reck-Peterson, R. L., A. Yildiz, A. P. Carter, A. Gennerich, N. Zhang, and R. D. Vale. 2006. Single molecule analysis of dynein processivity and stepping behavior. *Cell.* 126:335-348.
7. De La Cruz, E. M., A. L. Wells, S. S. Rosenfeld, E. M. Ostap, and H. L. Sweeney. 1999. The kinetic mechanism of myosin V. *Proc. Natl. Acad. Sci. USA.* 96:13726-13731.
8. Kodera, N., D. Yamamoto, R. Ishikawa, and T. Ando. 2012. Video imaging of walking myosin V by high-speed atomic force microscopy. *Nature.* 468:72–76.
9. Rosenfeld, S. S., and H. L. Sweeney. 2004. A model of myosin V processivity enzyme catalysis and regulation. *J. Biol. Chem.* 279:40100-40111.
10. Prigogine, I. 1978 Time, structure, and fluctuations. *Science.* 201:277-285.
11. Zot, H. G., J. E. Hasbun, and N. V. Minh. 2009. Striated muscle regulation of isometric tension by multiple equilibria. *PLoS ONE.* 4: e8052.
12. Zot, H. G., J. E. Hasbun, and N. V. Minh. 2012. Second chance signal transduction explains cooperative flagellar switching. *PLoS ONE.* 7: e41098.
13. DeWitt, M. A., A. Y. Chang, P. A. Combs, and A. Yildiz. 2012. Cytoplasmic dynein moves through uncoordinated stepping of the AAA+ ring domains. *Science.* 335:221-225.
14. Sowa, Y., and R. M. Berry. 2008. Bacterial flagellar motor. *Q. Rev. Biophys.* 41:103–132.
15. Terashima, H., S. Kojima, and M. Homma. 2008. Flagellar motility in bacteria: structure and function of flagellar motor. *Int. Rev. Cell Mol. Biol.* 270:39–85.
16. Block, S. M., J. E. Segall, and H. C. Berg. 1982. Impulse responses in bacterial chemotaxis. *Cell.* 31:215-226.
17. Kuo, S. C., and D. E. Koshland, Jr. 1989. Multiple kinetic states for the flagellar motor switch. *J. Bact.* 171:6279-6287.
18. Cluzel, P., M. Surette, and S. Leibler. 2000. An ultrasensitive bacterial motor revealed by monitoring signalling proteins in single cells. *Science.* 287:1652-1655.
19. Sourjik, V., and H. C. Berg. 2002. Binding of the Escherichia coli response regulator CheY to its target measured in vivo by fluorescence resonance energy transfer. *Proc. Natl. Acad. Sci. USA.* 99:12669–12674.
20. Sagi, Y., S. Khan, and M. Eisenbach. 2003. Binding of the chemotaxis response regulator CheY to the isolated, intact switch complex of the bacterial flagellar motor: lack of cooperativity. *J. Biol. Chem.* 278:25867–25871.
21. Duke, T. A. J., N. Le Novère, and D. Bray. 2001. Conformational spread in a ring of proteins: a stochastic approach to allostery. *J. Mol. Biol.* 308:541–553.

22. Bai, F., R. W. Branch, D. V. Nicolau, Jr, T. Pilizota1, B. C. Steel, P. K. Maini, and R. M. Berry. 2010. Conformational spread as a mechanism for cooperativity in the bacterial flagellar switch. *Science*. 327:685–689.
23. Lele, P. P., B. G. Hosu, and H. C. Berg. 2013. Dynamics of mechanosensing in the bacterial flagellar motor. *Proc. Natl. Acad. Sci. USA.* 110:11839–11844.
24. Tipping, M. J., N. J. Delalez, R. Lim, R. M. Berry, and J. P. Armitage. 2013. Load-dependent assembly of the bacterial flagellar motor. *mBio.* 4:e00551-13.
25. Wang, F., J. Yuana, and H. C. Berg. 2014. Switching dynamics of the bacterial flagellar motor near zero load. *Proc. Natl. Acad. Sci. USA*. 111:15752–15755.
26. Suzuki, T., K. Tanaka, C. Wakabayashi, E. Saita, and M. Yoshida. 2014. Chemomechanical coupling of human mitochondrial F1-ATPase motor. *Nat. Chem. Biol.* 10:930-6.

**Figure Legends:**

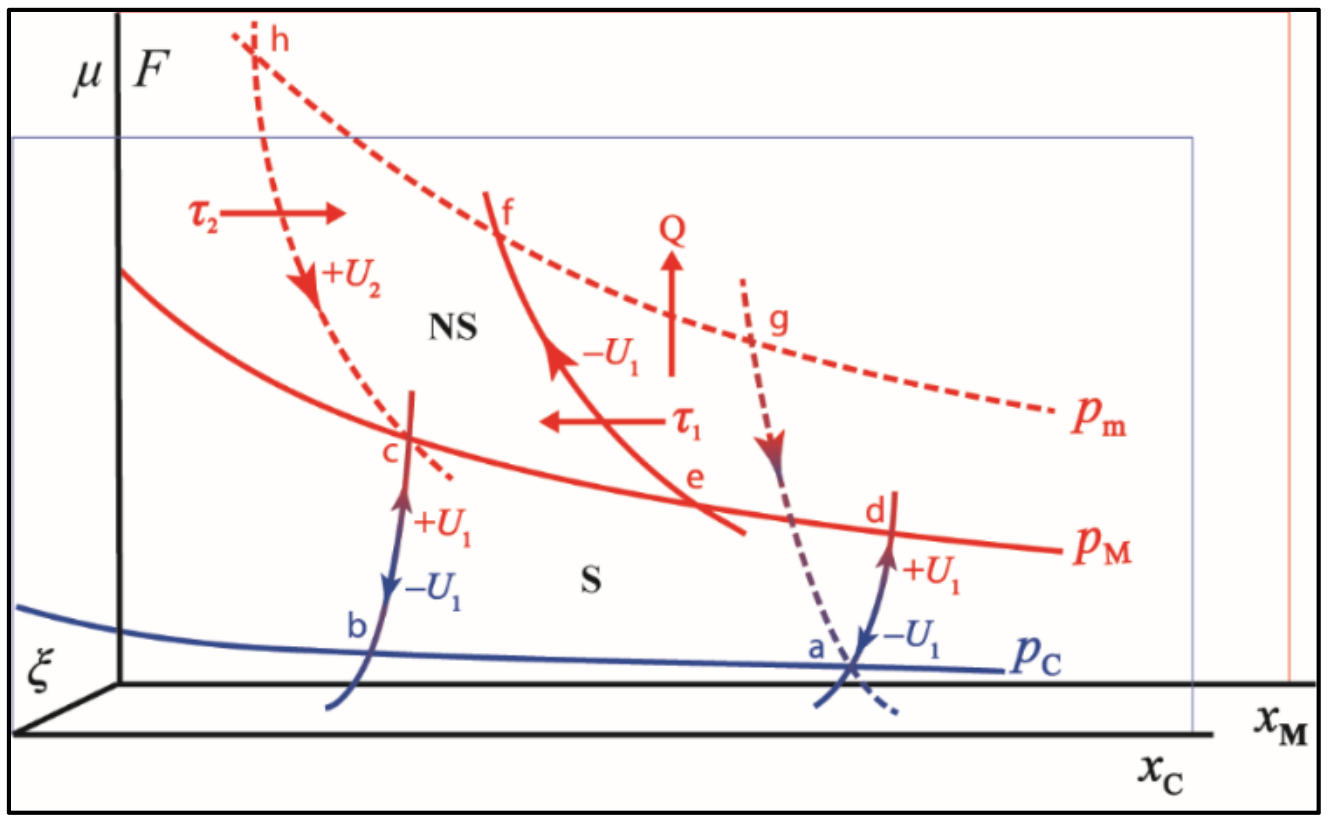

FIGURE 1. Conceptual state space of a kinetic recursive cycle (*cefh*) coupled to a chemical equilibrium (*abcd*). Free energy potentials of states (*C*, *M*) and kinetic intermediate (*m*) are diagramed on orthogonal planes (blue, red) for chemical ($\mu$•$d\xi$) and kinetic ($F$•$dx$) work. Adiabats, (*bc*, *ad*, *ga*) and accelerations (*ef*, *hc*) intersect instantaneous momenta ($p_C$, $p_M$, $p_m$). Interaction sites ($U_1$) are exchanged by mass action (*abc*, *adc*) and separated by kinetic energy (*cef*). Interactions that replace $U_1$ ($U_2$) are reconstituted by kinetic energy (*fhc*). External processes ($\tau_1$ and $\tau_2$) transfer kinetic energy isometrically to the system and the system evolves heat (Q) isothermally (*efga*). Spontaneous (S) and nonspontaneous (NS) refer to pathways *abcd* and *cefh*, respectively.

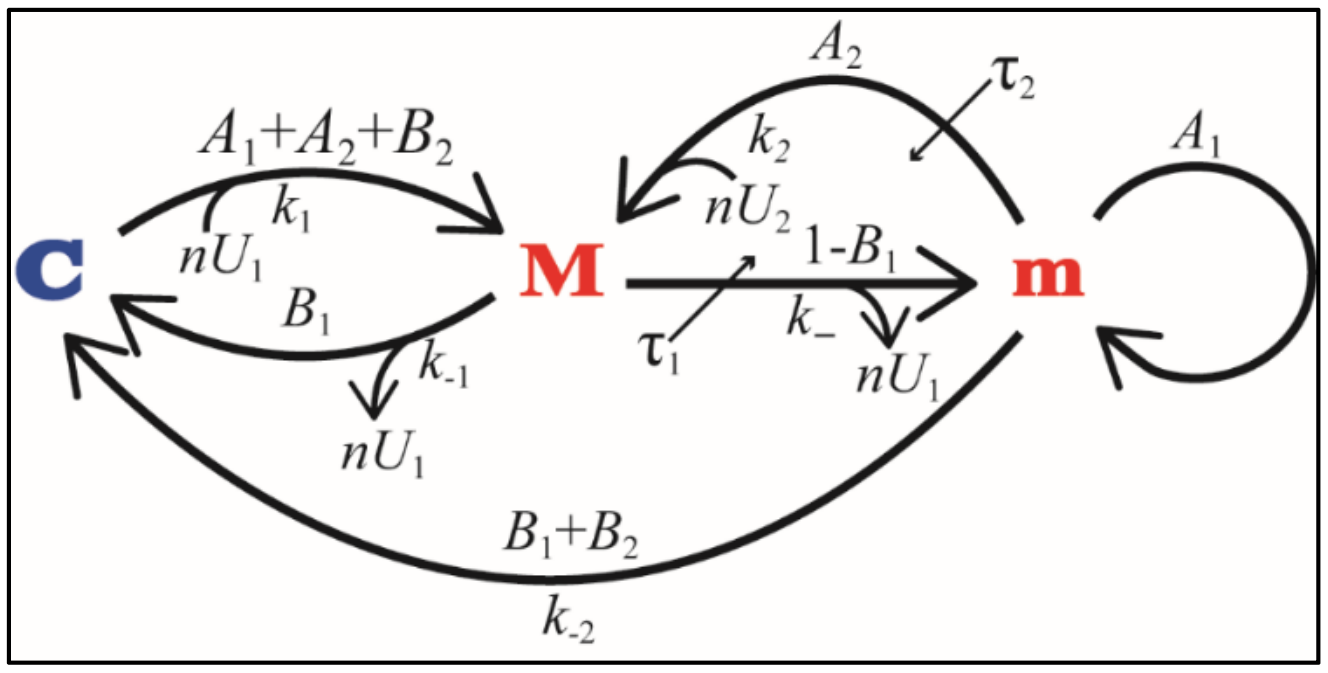

FIGURE 2. A system with a dynamic recursive cycle is represented as a steady-state and as a Markov chain. An equilibrium between states *C* and *M* having ensemble ($n \geq 1$) interactions ($nU_1$) is coupled to external processes ($\tau_1$ and $\tau_2$). Formation of kinetic intermediate (*m*) isolates $nU_1$. Reconstitution with interactions ($nU_2$) regenerates *M*. Rates of spontaneous processes ($k_1$, $k_{-1}$, $k_{-2}$) and non-spontaneous processes ($k_2$, $k_$) determine static probabilities of the Markov chain ($A_1$, $A_2$, $B_1$, $B_2$; Eq. 3-6).

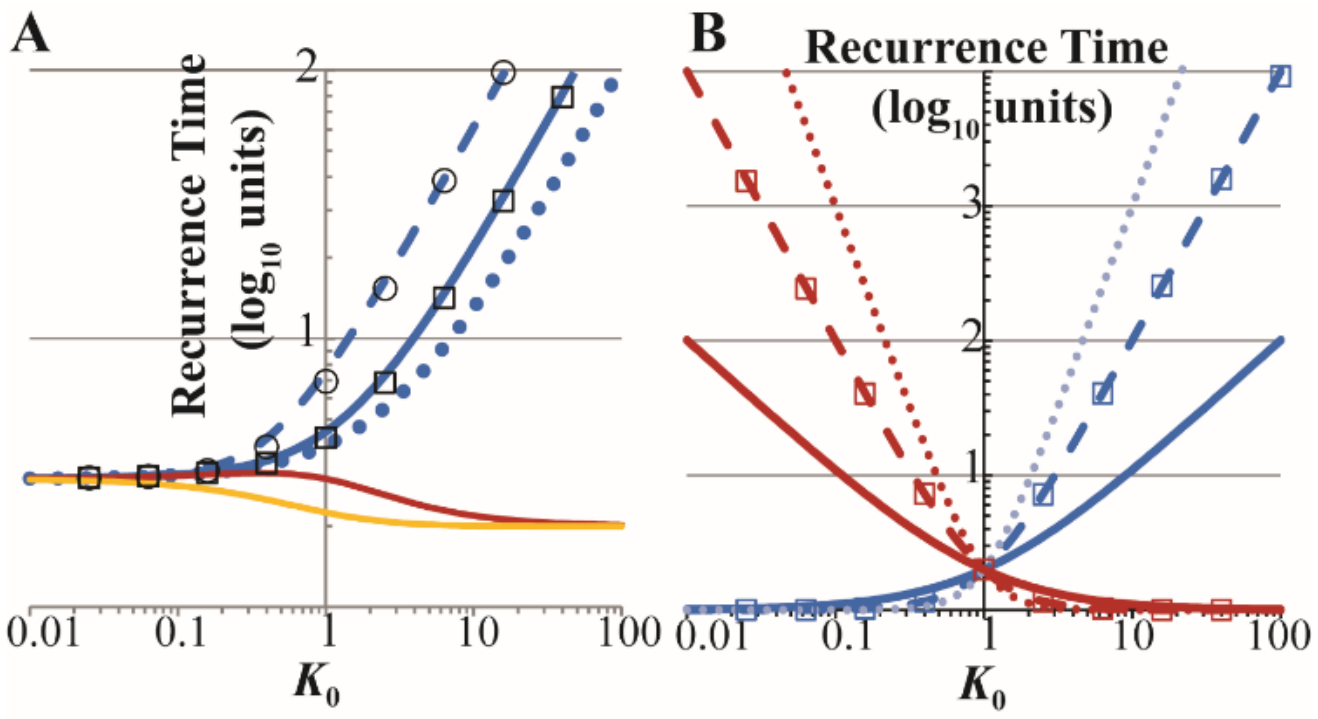

FIGURE 3. Recurrence times as a function of $K_0$. Discrete recurrence time for Markov states *C* (blue), *M* (red, **A**), *m* (gold, **A**), and either *M* or *m* (*Mm*) (red, **B**) are determined from stationary probabilities of matrix **J** (Supporting Material). In **A**, T1 conditions ($k_{-1} = 0$) are shown. With $n = 1$, $K_0$ is varied as a function of $k_{-2}$ and $\alpha = 0.5$ (dot), $\alpha = 1.0$ (solid, all colors), and $\alpha = 3.0$ (dash). For comparison, points corresponding to mean recurrence times for *C* by random walks ($10^5$ - $10^6$ steps; Fig. S1) are shown for $\alpha = 1$ (open square) and $\alpha = 3$ (open circle). In **B**, T1 conditions ($k_{-2} = 0$) are shown. Holding $\alpha = 1$, $K_0$ is varied as a function of $k_{-1}$ and $n = 1.0$ (solid line), $n = 2.0$ (dashed line), and $n = 3.0$ (dotted line). For comparison with $n = 2$, mean recurrence times are plotted for *C* (blue open square) and *Mm* (red open square) by random walks ($10^5$ - $10^6$ steps, Fig. S1).

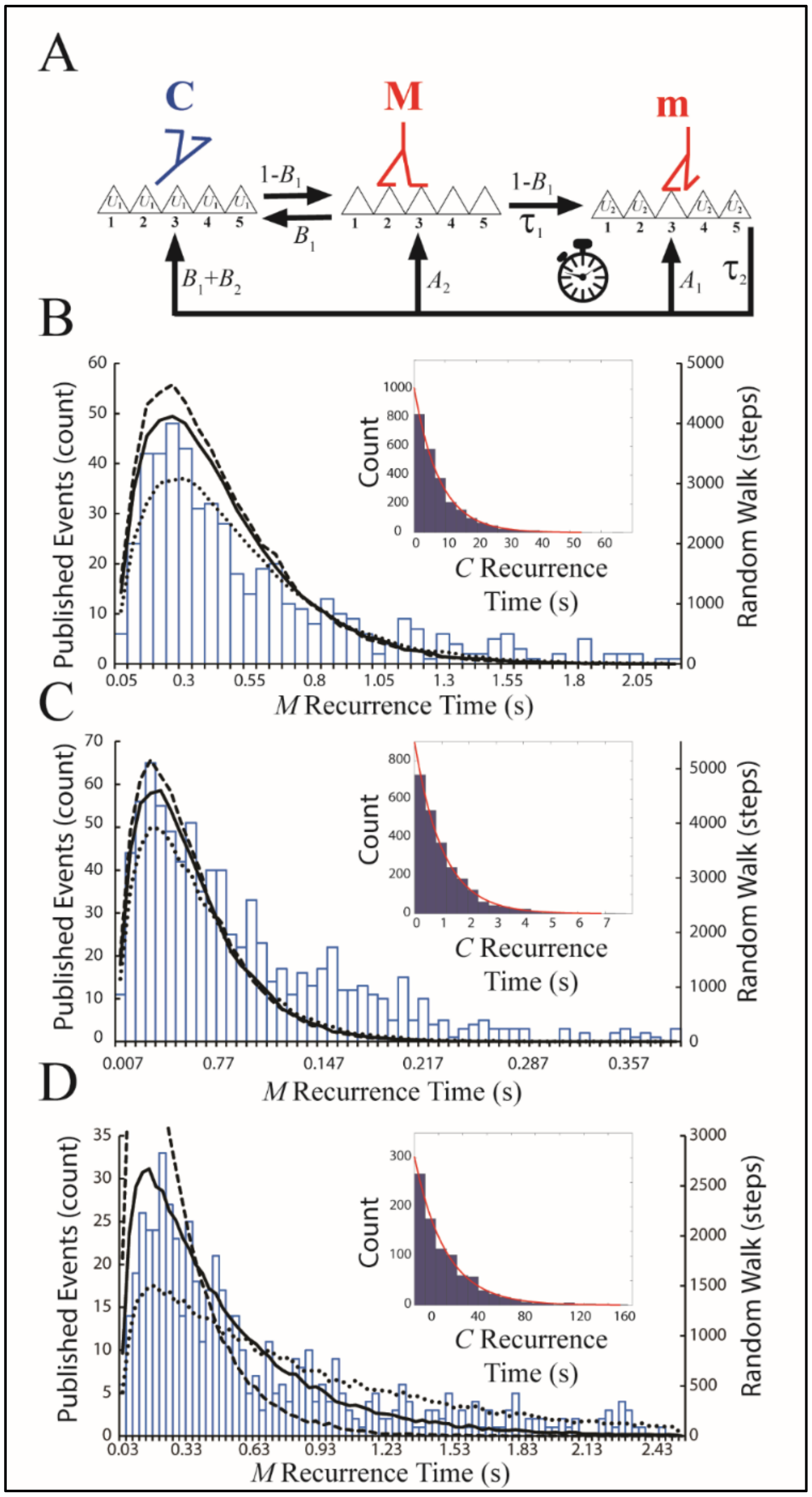

FIGURE 4. Application of *SCM* to a walking motor molecule. Panel **A.** Probabilities ($A_1$, $A_2$, $B_1$, $B_2$; Eq. 3-6) are shown for Markov states, *C*, *M* and *m* corresponding to detached (blue), thermodynamically attached (red M), and kinetically attached (red m), respectively. Markov steps have waiting times, $\tau_1$ and $\tau_2$, associated with the dynamic recursive cycle (clock). Free subunits, $U_1$ and $U_2$, are potential interactions for molecules in states *C* and *m*, respectively. Open bars reproduce data measured for MyoV in 2 μM ATP (**B**) and 2 mM ATP (**C**) and for dynein (**D**), respectively (5, 13). Lines represent recurrence times of *M* generated by random walk (Fig. S1), given constant conditions (Table 1) and varying one parameter: **B.** $K_0 = 5$ (dotted), $K_0 = 20$ (solid), and $K_0 = 200$ (dashed); **C.** $\alpha = 0.5$ (dotted), $\alpha = 1.0$ (solid), and $\alpha = 3.0$ (dashed); **D.** $\tau_2^{-1} = \tau_1^{-1}/16$ (dotted), $\tau_2^{-1} = \tau_1^{-1}/8$ (solid), and $\tau_2^{-1} = \tau_1^{-1}/4$ (dashed). Insets **B-D.**

Recurrence times of $C$ (filled bars) are generated with the random walks above for $K_0 = 20$ (**B**), $\alpha = 1.0$ (**C**), and $\tau_2^{-1} = \tau_1^{-1}/8$ (**D**) and fit to a single exponential (red).

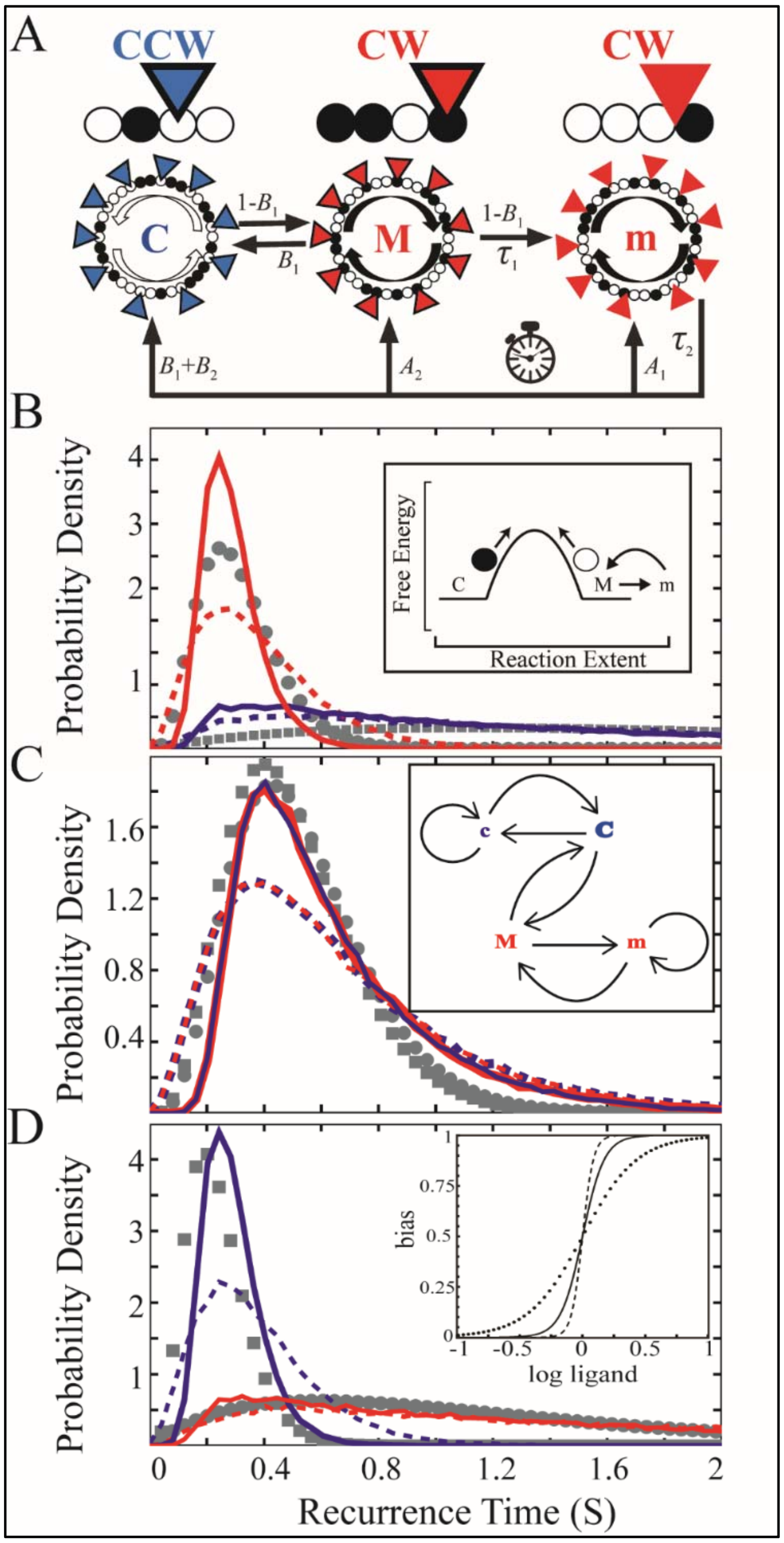

FIGURE 5. Application of *SCM* to bacterial flagellar switching mechanism. **A.** As the rotor travels in a counter clockwise (CCW) or clockwise (CW) direction, subunits (circles) that are ligand-free (open) or ligand-bound (closed) interact with the switch structure of stationary motor units (triangles). The angle of force (blue, red) is associated with thermodynamic states ($C$, $M$) and dynamic intermediate ($m$) of each motor unit ($n = 10$). Waiting times for each step of the recursive cycle (clock) are the inverse rate of the rotor ($\tau_1^{-1} = \tau_2^{-1}$). Probabilities ($A_1$, $A_2$, $B_1$, $B_2$) are a function of $M$ bias (Eq. 8) given rates in Table 1. **B-D.** Published probability density

distributions of intervals of CCW (gray squares) and CW (gray circles) rotations (1) are reproduced for CW bias equal to 0.1 (**B**), 0.5 (**C**), and 0.9 (**D**). Given each CW bias, the probability density distributions of recurrences times of *C* (blue) and *Mm* (red) are results of a random walk (Fig. S1) using probabilities determined by the corresponding *M* bias for $n = 2$ (dashed) and $n = 10$ (solid). Inset **B.** Example of a recursive cycle coupled to an equilibrium between switching states. Inset **C**. Example of a four-state Markov chain with recursive cycles for regenerating both *C* and *M*. Inset **D.** Ligand dependent *M* bias (Eq. 8) is shown for *n* equal 2 (dot), 5 (solid), and 10 (dash) motor units.

**Table 1**

Constants Used in Random Walk

| Experimental Fit | $n$ | $k_1$ | $k_{-1}$ | $k_k_{-2}$ | $\alpha$ | [a]$K_0$ or $K'_0$ | $\tau_1^{-1}$ | $\tau_2^{-1}$ |
|---|---|---|---|---|---|---|---|---|
| MyoV 2 mM ATP | 1 | 20 | 0 | 1 | 0.5-3 | 20 | 40 | 40 |
| MyoV 2 μM ATP | 1 | 5-200 | 0 | 1 | 1 | 5-200 | 5 | 5 |
| Dynein | 1 | 50 | 0 | 1 | 1 | 50 | 20 | 1.25-5 |
| Flagellar Switch | 2 | 1/9, 1, 9 | 1 | 0 | 1 | 1/9, 1, 9 | 8 | 8 |
| Flagellar Switch | 10 | 1/9, 1, 9 | 1 | 0 | 1 | 1/9, 1, 9 | 20 | 20 |

[a]Eq. 7, 8

**SUPPLEMENTAL MATERIALS**

Mean recurrence times (discrete) from matrix **J**. The time-invariant probabilities of *C*, *M* and *m*, i.e. stationary probability distribution, are elements, $i = 1, 2, 3$, respectively, associated with vector $\boldsymbol{\pi}$. This vector is proportional to the only eigenvector of matrix **J** that has unit eigenvalue. Mean recurrence times corresponding to *C*, *M* and *m* are proportional to $1/\boldsymbol{\pi}(i)$.

Q matrix (continuous time Markov). The Markov process becomes continuous by assigning a time for each step of the discrete process. The duration of each step is given by the waiting times, $\tau_1$ and $\tau_2$, of the $\tau_1$ and $\tau_2$ processes having Poisson rates, $\tau_1^{-1}$ and $\tau_2^{-1}$, respectively. Given the holding times, the probabilities of states *C*, *M*, and *m* for a given elapsed time are given by a continuous time matrix **Q** based on matrix **J**. Matrix

$$\mathbf{Q} = \begin{bmatrix} -\tau_1^{-1}(1-B_1) & \tau_1^{-1}(1-B_1) & 0 \\ \tau_1^{-1}B_1 & -\tau_1^{-1} & \tau_1^{-1}(1-B_1) \\ \tau_2^{-1}(B_1+B_2) & \tau_2^{-1}A_2 & -\tau_2^{-1}(1-A_1) \end{bmatrix}$$

where the decay to P(*C*|*C*), P(*M*|*M*), and P(*m*|*m*) are along the diagonal, and the rates across a row sum to zero. Stationary probabilities of the three Markov states (Fig. 1) are found in the null space of matrix **Q**. If $\tau_1^{-1}$ and $\tau_2^{-1}$ have the same value, the stationary probabilities of matrices **J** and **Q** are the same.

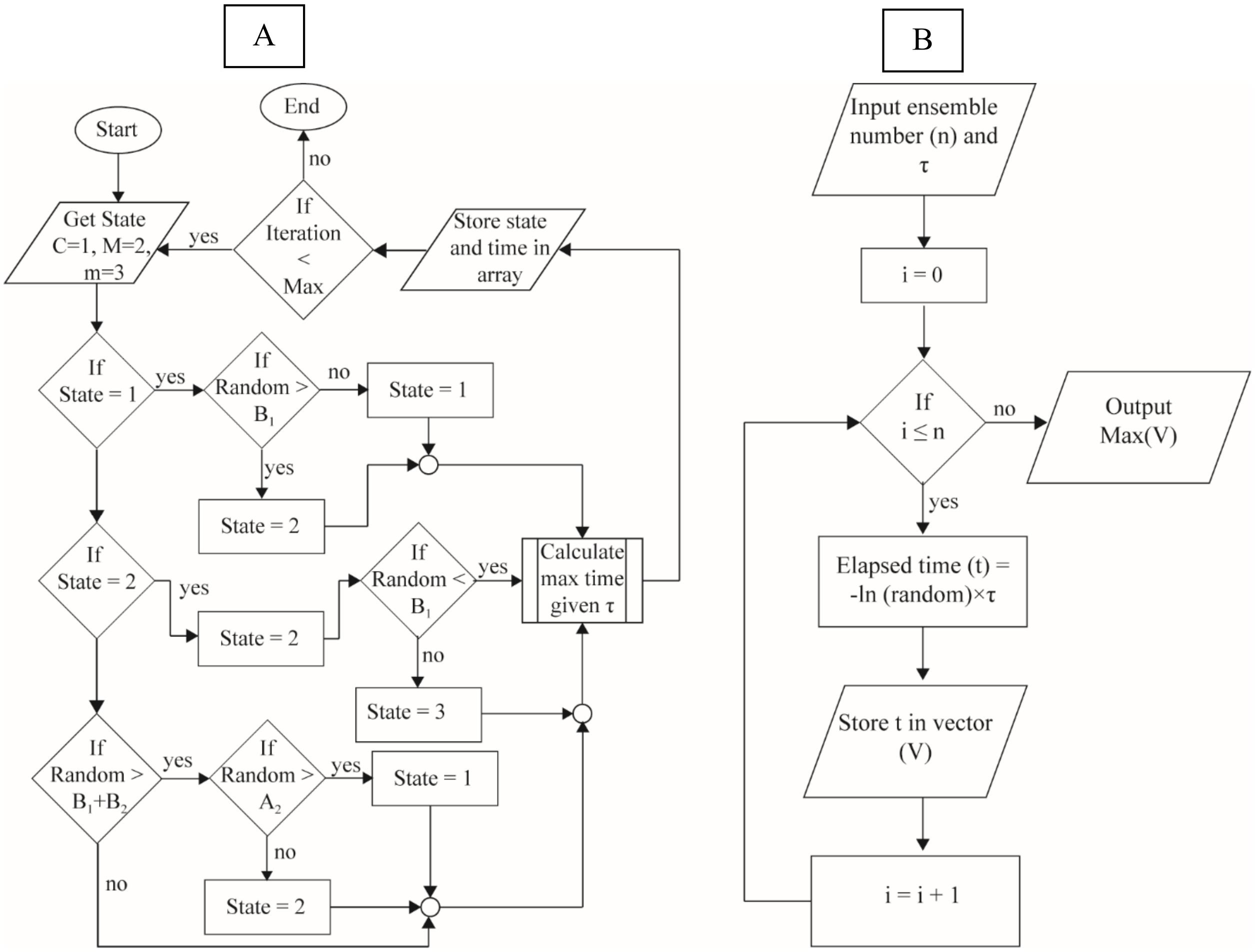

FIGURE S1. Flow charts of algorithms used to simulate a random walk of *SCM*. (A) Shown is an iterative loop of logic gates that evaluate transition probabilities (Fig. 2) with pseudorandom numbers. The maximum elapsed time of each step is longest continuous time event calculated with a routine (B). (B) This routine simulates the time ($t$) of a random decay from $P = e^{-t/\tau}$, where $P$ is a pseudorandom number and returns the maximum $t$ of $n$ trials, where $n$ corresponds to the ensemble number (Eq. 7, 8).